\documentclass[pre, twocolumn, amsmath, amssymb, superscriptaddress]{revtex4}

\usepackage{graphicx}
\usepackage{dcolumn}
\usepackage{bm}
\usepackage{braket}
\usepackage{xcolor}

\usepackage{amsmath}

\newcommand{\beq}{\begin{equation}}
\newcommand{\eeq}{\end{equation}}

\begin{document}

\title{ Determining  cancer cells division strategy}


\author{Mattia Miotto  \footnote{\label{corr} For correspondence write to: mattia.miotto@roma1.infn.it}}
\affiliation{Center for Life Nano \& Neuro Science, Istituto Italiano di Tecnologia, Viale Regina Elena 291,  00161, Rome, Italy}
\affiliation{Department of Physics, Sapienza University, Piazzale Aldo Moro 5, 00185, Rome, Italy}

\author{Simone Scalise}
\affiliation{Center for Life Nano \& Neuro Science, Istituto Italiano di Tecnologia, Viale Regina Elena 291,  00161, Rome, Italy}

\affiliation{Department of Physics, Sapienza University, Piazzale Aldo Moro 5, 00185, Rome, Italy}

\author{Marco Leonetti}

\affiliation{Center for Life Nano \& Neuro Science, Istituto Italiano di Tecnologia, Viale Regina Elena 291,  00161, Rome, Italy}
\affiliation{Soft and Living Matter Laboratory, Institute of Nanotechnology, Consiglio Nazionale delle Ricerche, 00185, Rome, Italy}
\affiliation{D-TAILS srl, 00161, Rome, Italy }

\author{Giancarlo Ruocco}
\affiliation{Center for Life Nano \& Neuro Science, Istituto Italiano di Tecnologia, Viale Regina Elena 291,  00161, Rome, Italy}

\affiliation{Department of Physics, Sapienza University, Piazzale Aldo Moro 5, 00185, Rome, Italy}

\author{Giovanna Peruzzi \footnote{\label{coaut} These authors contributed equally to the present work.}}
\affiliation{Center for Life Nano \& Neuro Science, Istituto Italiano di Tecnologia, Viale Regina Elena 291,  00161, Rome, Italy}

\author{Giorgio Gosti\ref{coaut}}
\affiliation{Center for Life Nano \& Neuro Science, Istituto Italiano di Tecnologia, Viale Regina Elena 291,  00161, Rome, Italy}
\affiliation{Soft and Living Matter Laboratory, Institute of Nanotechnology, Consiglio Nazionale delle Ricerche, 00185, Rome, Italy}

\begin{abstract}
Heterogeneity in the size distribution of cancer cell populations has been recently linked to drug resistance and invasiveness.  
However, despite many progresses have been made in understanding how such heterogeneous size distributions arise in fast-proliferating cell types  -like bacteria and yeast-,  comprehensive investigations on cancer cell populations are still lacking mainly due to the difficulties of monitoring the proliferation of the time scales typical of mammalian cells. 
From a reductionist cell dynamics point of view, the strategies allowing size homeostasis are roughly grouped into three classes, \emph{i.e.} timer, sizer, or adder. These strategies are empirically distinguishable given the phenomenological measurable relationship between the cell size at birth and at division, which requires following the proliferation at the single-cell level.
 Here, we show how it is possible to infer the growth regime and division strategy of leukemia cell populations using live cell fluorescence labeling and flow cytometry in combination with a quantitative analytical model where both cell growth and division rates depend on powers of the cell size.
 Using our novel approach,  we found that the dynamics of the size distribution of leukemia Jurkat T-cells is quantitatively reproduced by (i) a sizer-like division strategy, with (ii) division times following an Erlang distribution given by the sum of at least three independent exponentially-distributed times and   (iii)  fluctuations up to 15\% of the inherited fraction of size at division with respect to the mother cell size.
Finally, we note that our experimental and theoretical apparatus can be easily extended to other cell types and environmental conditions, allowing for a comprehensive characterization of the growth and division model different cells can adopt.
\end{abstract}

\maketitle

\section{Introduction}

Heterogeneity in cancer cell phenotypes enhances the population's capability of resisting fluctuating adverse environments~\cite{Kussell_2005, McGranahan2017, expexp}, administrations of drugs and anticancer treatments~\cite{Korolev2014}. In particular, recent works suggest that size heterogeneity specifically impacts cancer fitness, although the exact relationship between size and disease is poorly understood, as much as the strategies cancer cells adopt to create and maintain such heterogeneity~\cite{Jones2023}.
At a mechanistic level, cell size is determined by the combination of homeostatic processes operating at different timescales. If on shorter timescales, size is set by the capability of cells to regulate processes like water entry, membrane trafficking, and protein synthesis; on longer timescales, size results from the interplay between cell growth and division~\cite{Cadart2019}. 
As a result, under steady-state conditions, isogenic populations tend to maintain cell sizes around typical~\cite{Ginzberg2015}, type-specific values, although significant cell-to-cell variability is often observed~\cite{MODI20172408}.
In this respect, the inherent stochastic nature of molecular processes reflects in fluctuations of the production/degradation of cellular constituents and noisy partition of cell content/volume during cell division~\cite{Huh2010, Huh2011, Peruzzi2021}, ultimately leading to lognormal distributions of cell size, ranging over an order of magnitude~\cite{Hatton2023}, and characterized by coefficients of variation (CVs) of 0.1-0.3~\cite{scotchman2021identification, Jones2023}.

\begin{figure*}
    \centering
    \includegraphics[width=\textwidth]{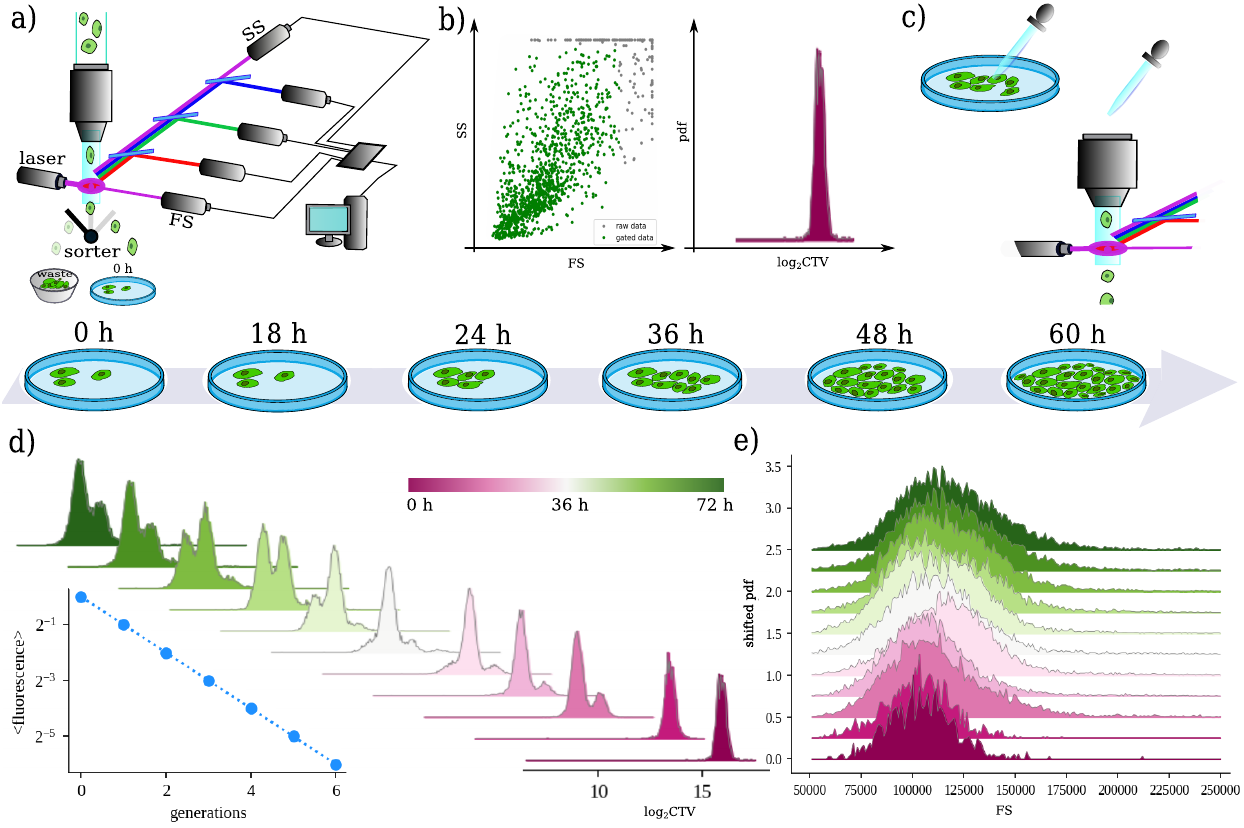}
    \caption{  \textbf{Experimental protocol.}  \textbf{a)}  Schematic representation of a fluorescence-activated cell sorter (FACS) working principle. Stained cells pass one at a time in front of a laser source that excites the fluorophores of the markers. The forward scattered (FSC), and side scattered (SSC) light, together with the one emitted by the dye fluorophores is collected and analyzed.  Eventually, a sorter divides cells according to certain thresholds on the measured intensities.
\textbf{b)} SSC vs FSC intensities for the initial population of marked cells together with the distribution of CTV (CellTrace Violet, cytoplasm marker) intensity. 
\textbf{c)} Schematic representation of the time course protocol: the initial sorted population is kept in culture (see Methods) and samples are collected at different time points and analyzed via a flow cytometer that collects both FSC, SSC, and CTV intensities for each analyzed cell.
\textbf{d)} Time evolution of the population CTV fluorescence intensity. Colors from purple to dark green represent different time points along the experimental time course,  from time zero to 72 hours. Inset shows the mean fluorescence as a function of the population generations.  
\textbf{e)} Density distribution of the forward scattering intensity of the cell population at different times. Colors range from purple to dark green as the time goes from zero (sorting of cells, i.e. start of the experiment) to 90 hours.}
    \label{fig:1}
\end{figure*}

Studies on bacteria and yeast populations showed that size distribution homeostasis can be obtained by modulating the amount of growth produced during the cell cycle in such a way that on average larger cells at birth grow less than small ones.  
 The strategies that cellular populations may adopt to reach and maintain such size homeostasis have been roughly classified into three distinct models, depending on whether divisions occur after a certain time (timer), upon reaching a certain size (sizer), or after the size of the cell has increased by a finite volume  (adder).

In particular,  a phenomenological linear relation (usually referred to as stochastic map) has been observed between the size at birth ($s_b$) and the size at division ($s_d$)~\cite{Phillips2019}: $s_d = a s_b + \eta $, where $\eta$ represent noise derived from the specific biological size control model, while the slope, $a$,  defines the size control models.  Usually, experiments look at the quantity, $\Delta = \braket{s_d} -\braket{s_b} = (a-1)\braket{s_b} + \braket{\eta}$. So that, for $a = 2$, the size at division is directly proportional to the size at birth; such the timer model hypothesizes that the cell size is controlled by a cell cycle timer that sets a time limit for the growth phase, and once the time limit is reached, the cell divides. 
Intuitively, if growth is exponential such a mechanism will end up in big cells to proliferate faster than small cells thus producing a divergent size variance. Only a linear growth regime under specific constraints on division symmetry is compatible with this homeostatic strategy~\cite{Thomas2018}. 
For $a=1$,  a certain volume is added which is uncorrelated to the initial size. In particular, the adder model proposes that the cell size increases by a constant amount during each cell cycle, regardless of the initial size. This behavior has been proposed for various populations of bacteria, cyanobacteria, and budding yeast populations \cite{Amir2014, Campos2014}. 
Finally, if $a=0$, the size at division is completely set by the stochastic term, which is called a sizer mechanism.
The sizer model suggests that the cell size is determined by a threshold size, and once the cell reaches this threshold size, it triggers the cell division process. A perfect sizer mechanism has been found for organisms like the fission yeast \textit{S.Pombe}\cite{Fantes1977, Pan2014}.

To determine which model is followed by the different kinds of cell types and determine the cell size distribution in both bacterial and eukaryotic cell populations, various experimental techniques have been developed~\cite{Cadart2014, EnricoBena2021}, comprising time-lapse microscopy~\cite{Wallden2016},  single-cell tracking~\cite{TaheriAraghi2015},  and gene tagging~\cite{Andersen1998}.
Parallel to experimental progress, different models of cell size regulation (with few key parameters) have been proposed.
From the pioneering works of Powell~\cite{POWELL1964} and Anderson~\cite{Anderson1969}, that compared analytical predictions with cell counts experiments to more recent works that proposed mathematical models to interpret both population and single cell-based data  \cite{Amir2014, Osella2014, 8619403, NietoAcuna2019,  evolat, tolomeo, Jia2021,  Lin2020}.

In particular, progresses in the quantitative measurement of single-cell features, using, for instance, mother machines, allow for the characterization of the growth and division strategy for fast-proliferating cell populations, like bacteria and yeasts (whose doubling times are on the order of hours). For other cell types, like mammalian ones,  it is harder to track several divisions due to lower division rates and more stringent growing conditions. 
Thus there is still not much understanding of size determination in cancer~\cite{Ho2018, Jones2023}.

In this paper, we propose a novel experimental protocol coupled with a minimal mathematical model to determine the homeostatic strategy adopted by populations of leukemia cells. In particular, the experimental protocol, we propose, makes use of flow cytometry measurements which yield information on the cell size (via the collected forward scattering signal~\cite{VargasGarcia2020}) and permit (via live cell fluorescent tags~\cite{Peruzzi2021}) the determination of cell lineage and partition noise. Experimental data are compared with the predictions of a minimal model where the variation of a single parameter allows the exploration of the different size-homeostasis strategies. 
In particular, we developed a quantitative analytical framework based on three key features: (i) cell growth and division rates depend on powers of the cell size,  (i) cell division can take place after the cell accomplishes a certain number of intermediate tasks, and (iii) daughter cells can inherit an uneven portion of the mother cell size.

We show that: (i) using forward scattering as a proxy for cell size~\cite{Li2015} allows to observe the dynamics of cell size distributions, which are in qualitative agreement with those shown by both numerical simulations of agent-based systems and a minimal analytical model based on a population balance equation. (ii) A simple exponential distribution of division times can not reproduce the observed dynamics, which instead requires an Erlang distribution with at least the sum of three independent exponentially distributed intermediate times. Finally, (iii) stratifying data according to cell generations allows us to fully infer the homeostatic strategy adopted by leukemia cells.  

Overall, our results provide insight into the mechanisms of cell size control and may contribute to the development of novel therapies for diseases that are characterized by abnormal cell size distribution.

\begin{figure*}
    \centering
    \includegraphics[width=\textwidth]{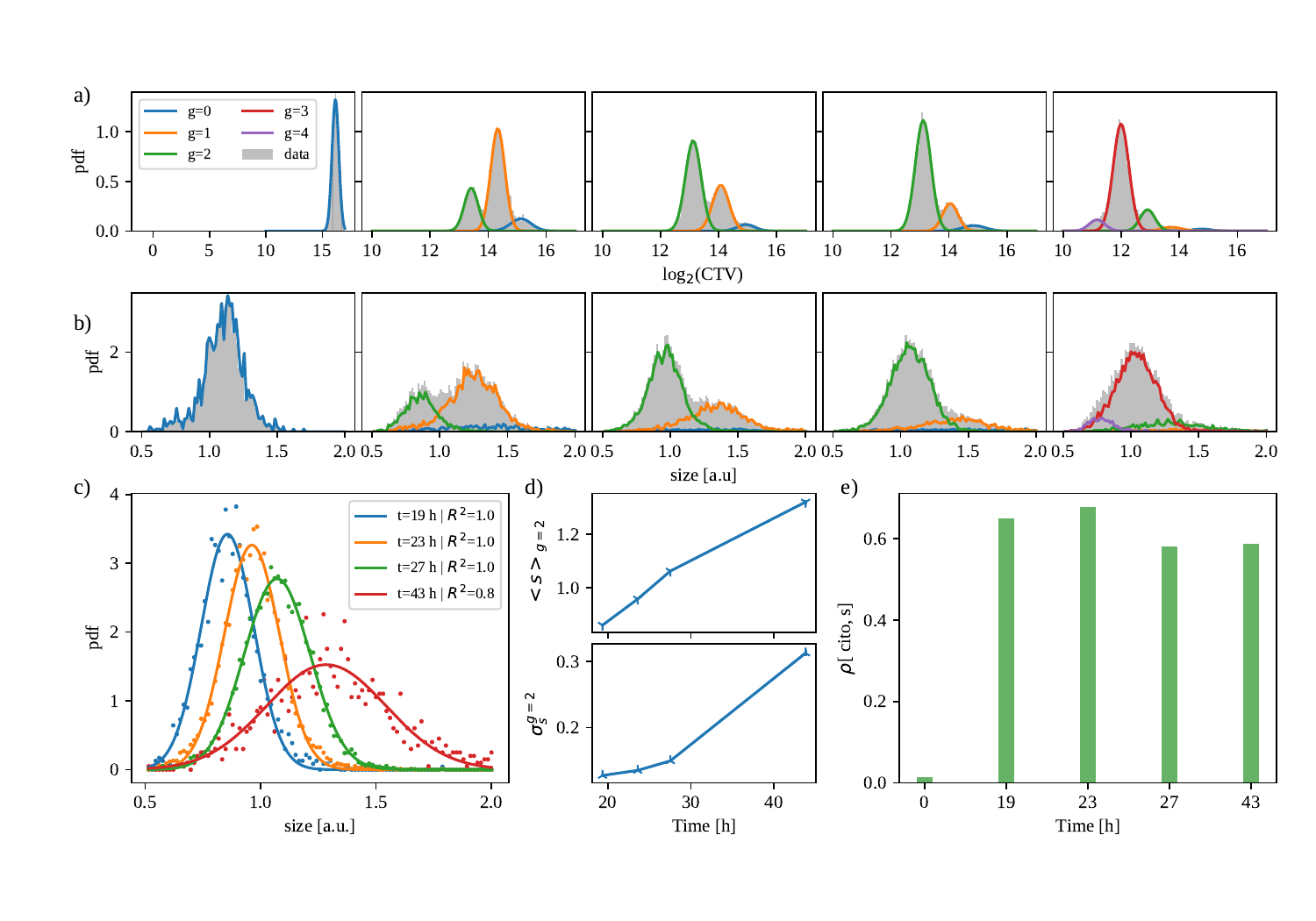}
    \caption{ \textbf{Analysis of forward scattering and CTV intensity data.} \textbf{a)} Density distribution (grey curves) and best fit of a Gaussian Mixture Model (colored curves) of the CTV fluorescence intensity measured in a Jurkat population at different times during its proliferation. From left to right, snapshots at 0, 19, 23, 27, and 43 hours from the CTV staining process.  Curves colored from blue to purple are ordered according to the identified generations.   
\textbf{b)} Same as in a) but for the measured FSC intensities. The colored curve highlights the subpopulations corresponding to the different generations that have been identified thanks to the Gaussian Mixture model fitting of the CTV intensities. Intensities have been rescaled by a factor of $10^5$.
\textbf{c)} Density distributions (dots) and best normal fit (lines) of the rescaled forward scattering intensity of the granddaughter's cells (second generation)  measured at different times during the proliferation of the population. The values of the $R^2$ for each fit are reported in the figure legend.
\textbf{d)} Mean and variance of the size of the grand-daughter subpopulation as a function of time. Size is quantified by the rescaled FSC intensity. Marker sizes are comparable with the Standard Error over the Mean and Variance of each time point.   
\textbf{e)} Pearson correlation coefficient of CTV and forward scattering intensities for the snapshots. 
    }
    \label{fig:3}
\end{figure*}

\section{Results}

\subsection{Experimental protocol}

To measure the cell size distribution and follow its dynamics, we developed a protocol based on flow cytometry measurements.
Extending the procedure we previously proposed to measure the partition noise of cellular compounds~\cite{Peruzzi2021}, we made use of CellTrace-Violet (CTV), a fluorescent dye, to mark cell cytoplasm (see Method section) and follow the proliferation of the population by looking at the dynamics of the fluorescence and forward scattering signals in time via a series of flow cytometry measurements.  As depicted in Figure~\ref{fig:1} and explained in more detail in the Methods, marked cells are first sorted (see panel a), so that an initial population with a narrow CTV distribution is selected  (Figure~\ref{fig:1}b).  The sorted population is then collected and cultured in standard growth conditions (see Methods). Samples of the population are then collected at different times, recording both CTV and forward scattering intensities, FSC, for each analyzed cell (see Figure~\ref{fig:1}c). 

Figure~\ref{fig:1}d shows the evolution of the distributions of $\log_2$ CTV fluorescence intensities. The  
purple curve (low-right corner of the panel) corresponds to the time zero, post-sorting distribution.
Looking at the CTV intensity at different times during the proliferation of the cell population, one observes a progressive shift of the initial fluorescence and the appearance of multiple peak distributions. As CTV homogeneously binds to cytoplasmic proteins, upon division, the fluorescence of each mother cell is divided into the two daughters as a result of the cell division process. 
Each division produces two daughters, thus on average the CTV distribution of the daughter cells has half the mean fluorescence of the mother distribution (see inset in Figure~\ref{fig:1}d).

Parallel to the evolution of the CTV intensity, we track the evolution of the FSC intensity. As it can be seen from Figure~\ref{fig:1}e, the distribution shifts toward higher values than those presented at the initial time point (purple curve), while its variance increases. This behavior can be explained by recalling that the initial population is sorted, consequently, its size distribution is out of equilibrium.



\subsection{Forward scattering evolution probes cell size dynamics}

To identify the different generations from the CTV intensity profiles, we applied a fitting protocol via a Gaussian Mixture Model of the form: $P(\ln x) = \sum_g w_g N(\ln x, \bar x_g, \sigma_g)$ combined with an Expectation Maximization algorithm (see Methods for details).  An example of the result of the fitting procedure is shown in  Figure~\ref{fig:3}a. From left to right, the distributions of the log2 of the CTV intensity for different time points are shown in grey, while the Gaussian distributions obtained as best fits of the experimental data are reported in different colors corresponding to the different identified generations.   
Together with the mean and variance of each Gaussian, the fitting procedure yields the probability of each cell to belong to the various identified generations.  
Since for each cell, both CTV and FSC signals are measured at each time point, we can use the information coming from the GM procedure to identify the subpopulations corresponding to different generations from the FSC distributions. Figure~\ref{fig:3}b displays the FSC intensity distributions of the same population from which the CTV distributions (shown in panel a) were measured. In this case, grey curves mark the total population, while colored ones correspond to the different generations. 
Comparing the distributions of the same generation in different snapshots, one notes that newer generations have distributions shifted toward smaller FSC values with respect to older ones. In particular, the distributions (dots) of the grand-daughter cells are reported in Figure~\ref{fig:3}c,  together with the best fits of a normal distribution for each distinct time point. 
Mean and variance as a function of the snapshot times are reported in Figure~\ref{fig:3}d.
Finally, we evaluate the Pearson correlation coefficients between CTV and FSC intensities as a function of time (Figure~\ref{fig:3}e). Results show that the correlation is null just after the sorting procedure and then reaches a high positive value (around 0.6) in the next snap times. This behavior appears reasonable since cytoplasm volume scales with the cell size; on the other hand, an almost zero correlation after sorting may indicate that the rates of CTV uptake are not strongly dependent on cell size.

\begin{figure*}
    \centering
    \includegraphics[width=0.99\textwidth]{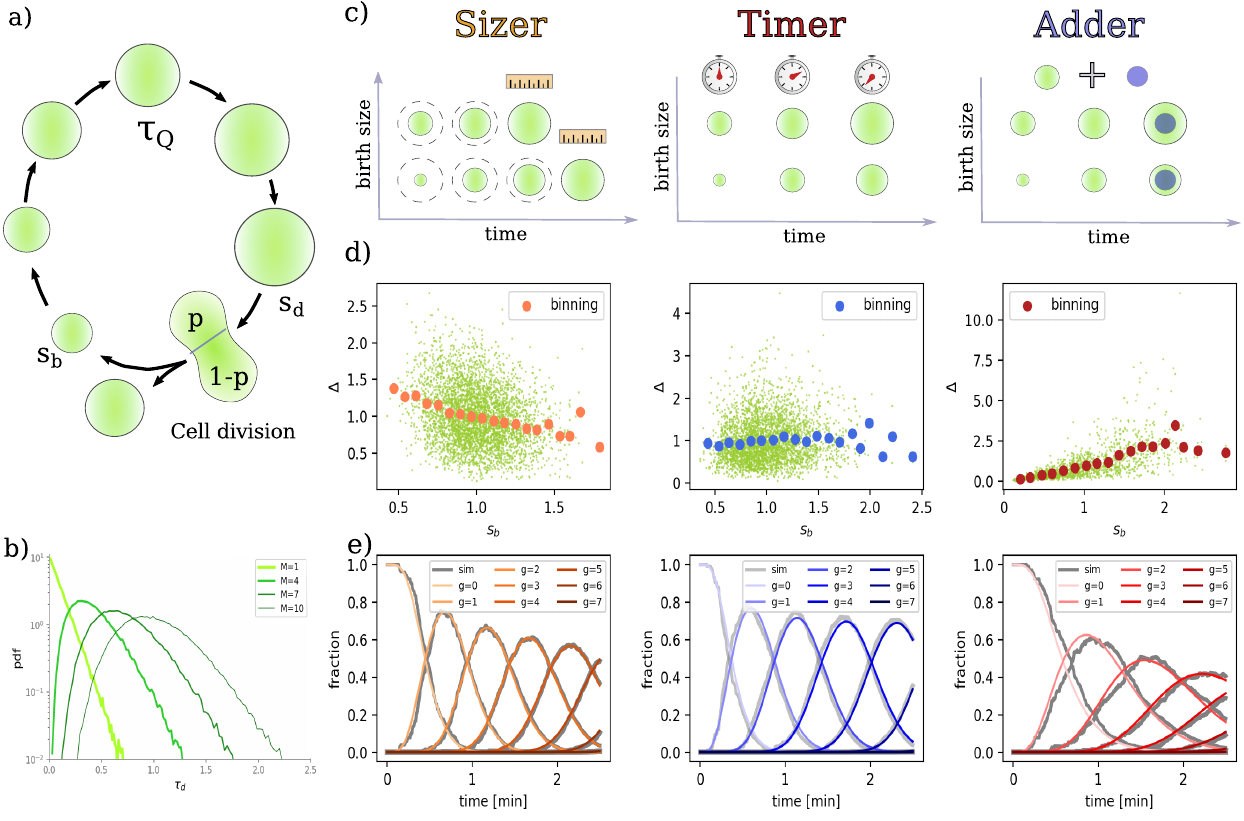}
    \caption{\textbf{Model of size-homeostasis.}
    \textbf{a)} Schematic representation of cell growth and division. A mother cell with starting size $s_b$ grows up to a size $s_d$ and then splits into two daughter cells whose starting sizes are fractions of the mother cell size. 
    \textbf{b)} Probability density distribution of the division times, $\tau_d$ as a function of the number of intermediate states a cell must visit before dividing. Time spent in each of the intermediate states is assumed to be exponentially distributed.
    \textbf{c)} From left to right, schematic representation of the sizer, timer, and adder mechanisms: cells grow (i) until a certain size is reached, for a certain time interval, or until a determined amount of size is added to the starting one. 
    \textbf{d)} Rescaled difference between size at division and birth, $\Delta = s_d-s_b$ vs rescaled birth size for the three size-homeostasis models. Both quantities are rescaled by the respective mean values. Light green dots represent the values obtained via a numerical simulation of a cell population in each regime, while orange, red, and blue dots a obtained binning over the x-axis. 
    \textbf{e)} Fraction of cells having divided $g$ times since the initial time of the simulation as a function of simulation time. From left to right, cells grow and divide according to a sizer, timer, or adder strategy, respectively.
    }
    \label{fig:2}
\end{figure*}

\subsection{Minimal  model for size dynamics}

At odds with single cell size measurements, where the homeostatic strategy can be inferred looking at the slope of the linear relation between birth and division sizes of the cells, 
 the interpretation of the data collected by our experimental protocol required a comparison with a  theoretical framework that accounts for the time evolution of the cell size distribution under different possible growth and division regimes, i.e. different size-homeostasis strategies.

\subsubsection{Exponentially distributed division times}
To allow for a comparison, we aimed at modeling the growth and division dynamics of a population of cells growing in a controlled environment. We assumed that each cell of the population is thus characterized by a size, $s$, which changes in time as the cell grows and divides and by its generation, $g$.

Referring to $n_g(s(t),t)$ as the number of cells in the population having size $s$  at time $t$ and having divided $g$ times, this number will evolve in time according to:

\begin{multline}
\label{eq:PBE}
\frac{\partial n_g(s,t)}{\partial t} + \frac{\partial \left(g(s) \cdot n_g(s,t)\right)}{\partial s} = -\gamma(s) n_g(s(t),t) +  \\ + 2 \int_0^{\infty} d\eta~ \gamma(\eta)~\phi(s|p\eta)~n_{g-1}(\eta,t)
\end{multline}

where $g(s)$ and $\gamma(s)$ are the size-dependent growth and division rate, respectively, and $\phi(x|py)$  quantifies the probability that a daughter cell inherits a fraction p of mother cell size y. See Appendix~\ref{app:PBE} for the derivation of Eq.~\ref{eq:PBE}. 

Along the lines of previous works~\cite{Osella2014, NietoAcuna2019, Totis2021, Jia2021}, we assumed that both growth and division rates are given by power of the cell size, i.e. we assumed $g(s) = \frac{ds}{dt}= \lambda s^\alpha$ and $\gamma(s) = \kappa s^\beta$. It can be easily shown that with this minimal assumption, it is possible to recover all main size-homeostatic strategies tuning the parameter $\omega = \beta -\alpha$ (see Appendix \ref{app:strategies} and \cite{Nieto2020}).

If we now look at the variation of the total number of cells at generation $g$, we have:

\beq
\dot N_g = \frac{d}{dt} N_g(t) =  \int_0^\infty \frac{\partial}{\partial t} n_g(s,t) ds 
\eeq

which can be recast as

\begin{multline}
\dot N_g = \int ds \big[ -\frac{\partial}{\partial s}\left( \lambda s^\alpha n_g(s,t)\right)  - k s^\beta  n_g(s,t) + \\ +  2k \int d\eta \eta^\beta \int ds \phi(s|p\eta)~ n_{g-1}(\eta,t) 
\end{multline}

The first term of the integral goes to zero thanks to the fact that either the size or the number of cells is zero in the integration extrema; thus the above equation becomes:

\beq
\label{eq:ndot_g}
\frac{\dot N_g}{N_g} = - k \braket{s^\beta}_g  +  2k \braket{s^\beta}_{g-1}  \frac{N_{g-1}}{N_g} = \Phi_g
\eeq

where $<x>_g = \int ds x \rho_g$ and we introduced the probability of finding a cell with size $s$ at time $t$ and generation $g$ as

\beq
\rho_g(s,t) = \frac{n_g(s,t)}{N_g(t)}.
\eeq

The fraction $N_{g-1}/N_g$ is difficult to be experimentally measured. Thus, we want to recast it in a more handy form. To do so, we define the fraction of cells belonging to a certain population at each time as:
\beq
P_g(t) = \frac{N_g(t)}{\sum_q N_q(t) }
\eeq

Eq.~\ref{eq:ndot_g} can be used to compute the dynamics of the fractions, $P_g$. In fact, we have that~\cite{expexp}:

\begin{multline}
\dot{P_g} = \left( \dot{ \frac{N_g(t)}{\sum_q N_q(t) }} \right) =  
\frac{\dot{N_g}}{\sum_q N_q} - P_g   \frac{\sum_q \dot{N_q}}{\sum_q N_q}
 \end{multline}
 
 Using Eq.~\ref{eq:ndot_g}, we can explicit $\dot{N_g}$ and obtain:
 
 \begin{multline}
 \label{eq:fractions}
 \dot{P_g} = 
 -k\braket{s^\beta}_g P_g + 2k\braket{s^\beta}_{g-1} P_{g-1} + \\ + P_g \sum_q  \left( k\braket{s^\beta}_q P_q - 2k\braket{s^\beta}_{q-1} P_{q-1} \right)
 \end{multline}

It remains to obtain expressions for the mean and variance dynamics. Again, we can start from Eq.~\ref{eq:ndot_g}.  In fact, 
\beq
\dot{n_g} = \dot{N_g} \rho + N_g \dot{\rho_g} 
\eeq

and 

\begin{multline}
- \frac{\partial}{\partial s}\left( \lambda s^\alpha n_g(s,t)\right) 
- k s^\beta  n_g(s,t) + \\ + 2 k \int d\eta (\eta)^\beta~\phi(s|p\eta)~n_{g-1}(\eta,t) =\\= \Phi_g N_g \rho_g + N_g\dot{\rho_g}~.
\end{multline}

Reordering and dividing by $N$, we get

\begin{multline}
\label{eq:rho_ev}
\dot{\rho_g} = - \Phi_g \rho_g 
- \frac{\partial}{\partial s}\left( \lambda s^\alpha \rho_g\right) 
- k s^\beta  \rho_g + \\ + 2 k \int d\eta \eta^\beta~\phi(s|p\eta) \rho_{g-1}(\eta,t) \frac{N_{g-1}}{N_{g}}~.
\end{multline}

Without loss of generality, one can express $\phi$ as

\beq
\label{eq:noise}
\phi(s|p\eta) = \int_0^1 dp~\pi(p)\delta(s-p\eta)
\eeq

where $\pi(p)$ is a general probability function of the fraction of inherited cell size.

Thanks to Eqs.~\ref{eq:rho_ev} and ~\ref{eq:noise}, we can easily compute the distribution moments evolution equations as:

\begin{multline}
\label{eq:moments}
\braket{\dot{s}^i}_g =  \lambda \cdot i\cdot \braket{s^{(\alpha + i -1)}}_g - \Phi_g\braket{s^i}_g  -k\braket{s^{(\beta + i)}}_g + \\+ 2~k\left< p^i \right>_\pi \braket{s^{(\beta+ i)}}_{g-1}  \frac{P_{g-1}}{P_{g}}
\end{multline}

where $<p^i>_\pi$ refers to the i-th moment of $\pi(p)$ (see Appendix~\ref{app:noise} for details on how the last term is obtained).

\subsubsection{Erlang distributed division times}

Assuming size-dependent growth and division rates end up producing exponentially-distributed division times~\cite{NietoAcuna2019}, while it has been previously shown how intergeneration division time statistics is better captured by Erlang distributions\cite{Yates2017, NietoAcuna2019}. 
To retrieve an Erlang-like distribution, we introduce a series of intermediate states, the cell has to go through, before starting the division as done for instance by Nieto \textit{et al.}~\cite{Nieto2020} to describe \textit{E. coli} size dynamics. Each of the state's duration has an exponential distribution of times.
  Introducing another index accounting for the intermediate states the cell transit in before division, and repeating all calculation (see Appendix ~\ref{app:checkpoints}), one ends up with: 

\begin{multline}
\label{eq:fractions_q}
\braket{\dot{s}^i}_{g,q} =  \lambda \cdot i\cdot \braket{s^{(\alpha + i -1)}}_{g,q} - \Phi_{g,q}\braket{s^i}_{g,q}  -k\braket{s^{(\beta + i)}}_{g,q} +  \\ + 
k\left< 2~p^i\right>_\pi^{\delta_{q,0}} \braket{s^{(\beta+ i)}}_{g,q -1}  \frac{P_{g, q -1}}{P_{g,q}}
\end{multline}

 where $\braket{\cdot}_{g,q}$ stands for the statistical average over $\rho_{g,q}$,   the density of cells that divided $g$ times and passed $q$-th out of Q intermediate states and $\delta_{i,j}$ is the Kronecker delta. Similarly, the fraction of cells being at generation $g$ and state $q$ evolves according to

\begin{multline}
\label{eq:moments_0}
\dot{P_{g,q}} =  -k\braket{s^\beta}_{g,q} P_{g,q} + 2k\braket{s^\beta}_{g,q-1} P_{g,q-1} + \\ + P_{g,q} \sum_{h,w}  \left( k\braket{s^\beta}_{h,w} P_{h,w} - 2^{\delta_{w,0}}k\braket{s^\beta}_{h,  w -1} P_{h, w -1} \right)
\end{multline}

for  $q > 0$, and  

\begin{multline}
\label{eq:moments_q}
\dot{P_{g,0}} =  -k\braket{s^\beta}_{g,0} P_{g,0} + k\braket{s^\beta}_{g-1,0} P_{g-1,0} + \\ +  P_{g,0} \sum_{h,w}  \left( k\braket{s^\beta}_{h,w} P_{h,w} - 2^{\delta_{w,0}}k\braket{s^\beta}_{h, w-1} P_{h, w-1} \right)  \end{multline}

otherwise.

Equations~\ref{eq:fractions_q} and ~\ref{eq:moments_0}, \ref{eq:moments_q} fully describe the dynamics of the cell population, however except for some specific sets of parameters, this set of equations is not closed; in fact, the time derivative of the i-th moment may contain higher moments depending on the values of $\alpha$ and $\beta$. 
Indeed, the set is closed only in the case of a division rate that does not depend on the cell size (i.e. $\beta=0$). To solve the system in the general case, we must choose a moment closure strategy. To do so, we exploit our findings on the FSC distributions stratified by generations~\cite{Totis2021}. In fact, as the latter are fitted by normal distributions, we assume that the single generation size distributions have normal moments (note that this is not the case of the total population size distribution, that shows a log-normal distribution instead)  opt for a normal moment closure (see Appendix \ref{app:moment_closure} for details). 

To validate the obtained relations and test the adopted moment closure, we compare the solution of the differential equations with the results of stochastic simulations of an agent based model, where an initial population of cells grow and divide following the same grow and division rates functional form used in Eq.~\ref{eq:PBE}. To associate at each cell a proper division time, a Gillespie procedure has been adopted. See the Method section for a detailed description of the stochastic simulation protocol. The outcomes of the simulations are recapitulated in Figure~\ref{fig:2}. 
In particular, Figure~\ref{fig:2}a provides a schematic representation of the life cycle of a single agent, i.e. cell in the population: a cell is born with initial size, $s_b$; it grows according to a certain growth rate, $g(s)$, for a certain set of times, ${\tau_q}_0^Q$, which encode a series of independent intermediate states the cell has visited before actual division (see appendix \ref{app:checkpoints}). From a biological point of view, such states can be linked to the phases of the cell cycle. Upon reaching the division size, $s_d$, the mother cell splits into the two daughter cells, one inheriting a fraction p of the mother volume and the other keeping the remaining $1-p$ fraction.

To begin with, we verified that such a framework reproduces the expected sizer, timer, and adder (see Figure~\ref{fig:2}c.d) behavior upon varying the $\gamma = \beta - \alpha$ parameter. Indeed, simulations with $\alpha=1$ and $\beta$ taking values 2, 0, or 1 produced the expected trend for the sizer, timer, and adder, respectively. Next, we compared the solution of the model, in the normal closure approximation, with the results of the agent-based stochastic simulations. In Figure~\ref{fig:2}e, we show the results for the fractions of cells found in different generations as a function of time. There is a perfect accord between model and simulations as testified by the values of the $R^2$ of ~0.9.
These results both confirm the analytical calculations and the choice of the moment closure for such kind of dynamical process.


\begin{figure*}
    \centering
    \includegraphics[width=0.95\textwidth]{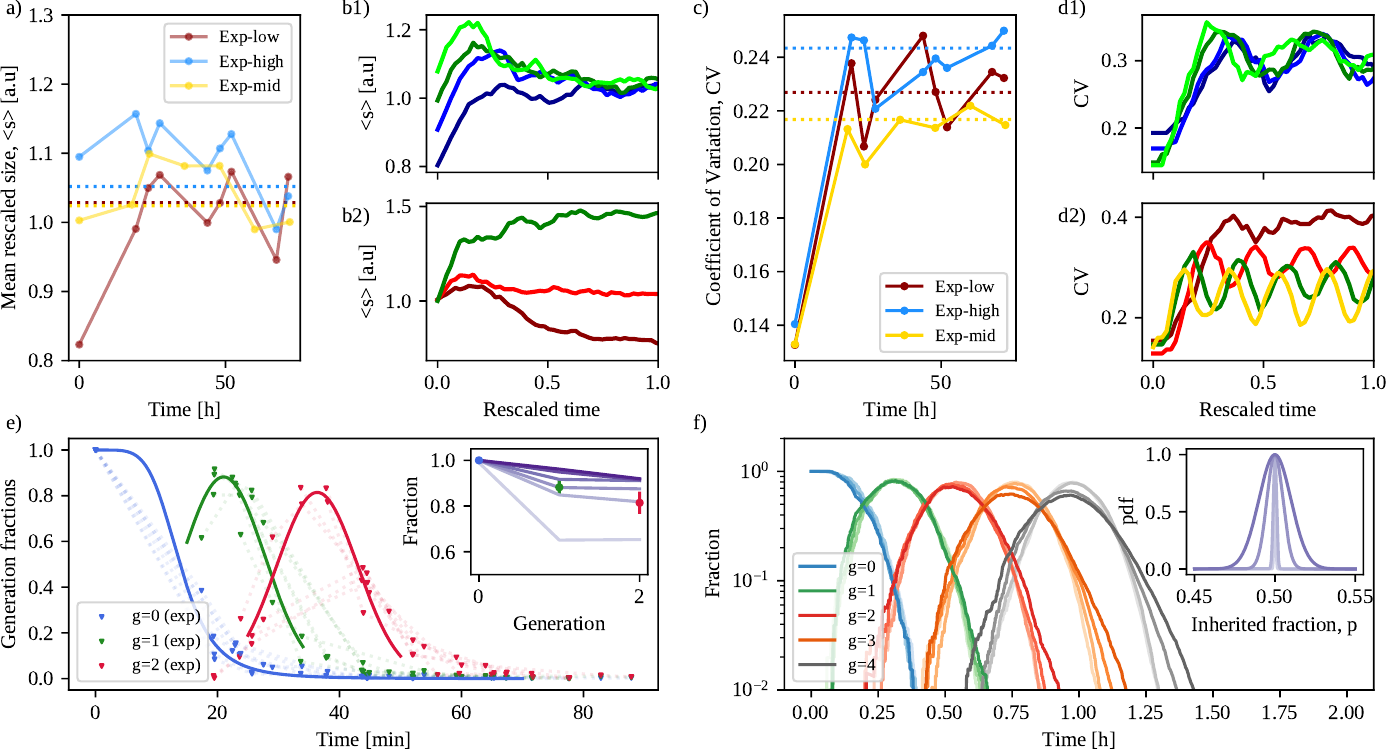}
    \caption{\textbf{Role of the model key parameters.}
    \textbf{a)} Mean of the rescaled size distribution, $<s>$ as a function of time for three Jurkat cell populations that have been sorted for low, medium, and high values of forward scattering intensity at time zero. Dotted horizontal lines mark the average values $<s>$ reaches after 2 days of proliferation.
 \textbf{b1)} Mean size of the population \, as a function of time obtained solving Eqs.~\ref{eq:moments} for different values of the mean size of the population at time zero, $<s(0)>$.  
\textbf{b2)} Same as in panel b1) but for different values of the ration $\lambda/\kappa$.
\textbf{c)} Same as in panel a) but for the coefficient of variation, CV.
\textbf{d1)} Same as in panel b1) but for the CV.
\textbf{d2)} Same as in panel d1) but varying the division rate exponent, $\beta$.
\textbf{e)}  Fraction of mother (blue), daughter (green), or granddaughter (red) subpopulations as a function of time. The maximum observed fractions of cells for the three different generations are reported in the figure inset.  Maximum values are computed fitting the mother fraction with a sigmoidal function, while daughter and gran-daughter ones are fitted with two normal distributions. Expected maximum fractions obtained solving Eqs.~\ref{eq:fractions_q} for different number of intermediate tasks, T, are shown in shades of purple. Maxima increases as a function of T.
\textbf{f)} Fraction of cells belonging to different generations as a function of time obtained solving equations~\ref{eq:fractions_q} for levels of noise in the cell size division between daughter cells.  Fraction of inherited volume, p, is described by a normal distribution centered in $1/2$ and with different variances as shown in the inset.  
}
    \label{fig:4}
\end{figure*}

\begin{figure*}
    \centering
    \includegraphics[width=\textwidth]{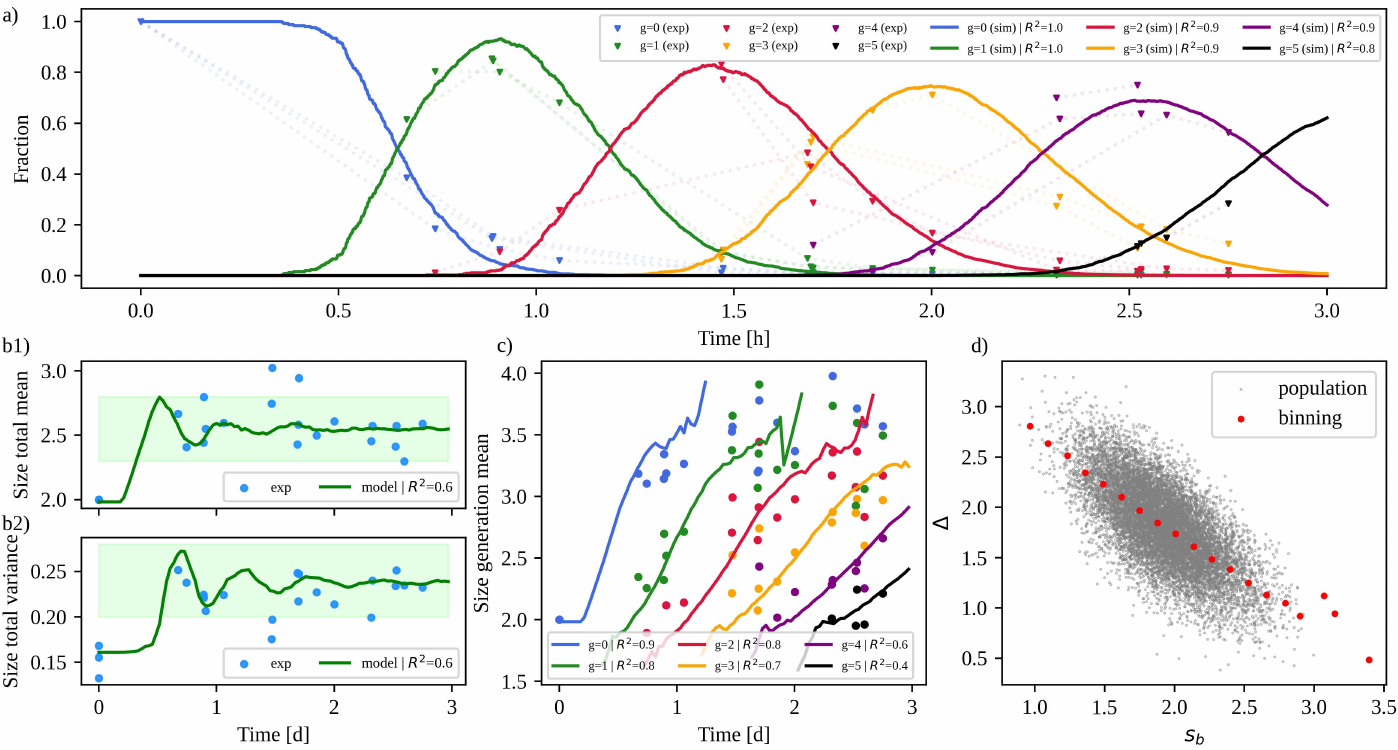}
    \caption{\textbf{Model vs experimental data.} \textbf{a)} Measured fractions of cells in different generations as a function of time (dots) and curves given by the best fit of the minimal model, described by Eqs.~\ref{eq:moments} and Eqs.~\ref{eq:fractions}.   \textbf{b1-b2)} Same as in panel a) but for the mean re-scaled total size and its CV given by the forward scattering measurements.    \textbf{c)} Mean rescaled size for different generations as a function of time (dots) and trends given by the best-fit solution of the model (lines).  \textbf{d)}  Rescaled difference between size at division and birth, $\Delta = s_d-s_b$ vs rescaled birth size for the best-fit solution of the model.  Grey dots represent the values obtained via a numerical simulation of a cell population, while red dots a obtained binning over the x-axis. 
    Best fit parameters are: $\lambda/\kappa=97.67$, $Q=5$, $<p>=0.5$, $\sigma^2_p=0.005$, $\alpha=1$, and $\beta=6$. }
    \label{fig:5}
\end{figure*}

\subsection{Model parameters govern distinct and measurable aspects of cell dynamics}

The derived framework depends on several parameters, thus, we next proceeded to characterize the role of the different model parameters and their effects on the quantities we can measure experimentally, i.e. the population size distribution moments, their per-generation stratifications, and the relative abundances of cells in different generations during the dynamics.

At first, we focused on the mean size of the whole population. Measuring the mean FSC as a function of time, we found that it reaches a well-defined oscillating non-equilibrium steady state~\cite{ecolat}, which does not depend on the initial value (see Figure~\ref{fig:4}a). Comparing the trend shown by experimental data with those of the model (Figure~\ref{fig:4}b1-b2),  we found that the asymptotic mean size is not a function of the starting mean sizes (see Figure~\ref{fig:4}b1) but it is modulated by the ratio of the rate coefficients, $\lambda / \kappa$. In particular, Figure~\ref{fig:4}b2 clearly shows that the higher the ratio $\lambda / \kappa$, the higher the population mean size in the long time limit.  Looking at the trend of the Coefficient of Variation, CV, instead, we found that the size of the Jurkat population displays a CV with an oscillating behavior around a value of about twenty-three percent as shown in Figure~\ref{fig:4}c. A comparison with the model trends again shows that this behavior is qualitatively reproduced by the model. Moreover, the key parameter modulating the variance of the cell size distribution is the exponent of the division rate, $\beta$. In particular, the higher the exponent the lower the fluctuations of the cell sizes (see Figure~\ref{fig:4}d2).

To explore the role of the remaining model parameters, i.e. the number of intermediate states and the division noise, we moved to consider the time evolution of the fractions of cells per generations.
Figure~\ref{fig:4}e shows the results for three timecourses 
as dots. As one can see, experimental data exhibit a trend qualitatively similar to those obtained solving Eqs. \ref{eq:fractions} and shown in Figure~\ref{fig:2}e.  Notably, the maximum fraction of observed first generation cells in the population is $0.85\pm0.05$. This value is not compatible with a single-task model of cell growth and division but requires a cell division time given by the sum of at least three independent exponentially distributed times. Indeed, this can be seen comparing the predicted maximum fraction of daughter cells obtained solving Eqs.\ref{eq:fractions}. The inset in Figure~\ref{fig:4}e displays the maximum fraction values obtained by changing the number of states, M, from one to 6.  
Finally, we quantified the role of division noise in terms of the shapes of the generation fractions curves. As discussed in the previous sections, given the high correlation between FSC and CTV intensity, we assume that the size at division follows the same statistics of the cytoplasmic components. In previous work, we found that Jurkat cells partition their cytoplasm symmetrically~\cite{Peruzzi2021}. Thus, we assumed that even the fraction of inherited volume is a random variable with a normal distribution, centered in $1/2$ and having a certain variance, $\sigma_p^2$ (see inset in Figure~\ref{fig:4}f). Solving the model equations with different values of $\sigma_p$ while keeping all other parameters fixed gives the trend reported in Figure~\ref{fig:4}f. It can be seen that the higher the level of division noise, the more the maximum fraction of cells per generation decreases, while the same generation tends to endure for longer times, i.e. smaller cells can be produced that require longer times to divide. Note that this reflects an increase in the total population variance.


\subsection{Size dynamics behaves according to a size-like homeostatic strategy.}

Finally, we compared the prediction of the model against the collected data. As discussed in the previous section, Eqs.~\ref{eq:moments} and Eqs.~\ref{eq:fractions} depend on four parameters governing the growth and division rates, two parameters fixing the first two moments of the initial size distribution, the number of intermediate tasks, Q,  and the variance of the inherited size fraction. In particular, mean and variance of the initial size distribution are directly measurable from the starting post-sorting forward scattering distribution, while the exponent of growth rate is fixed to 1, as required to reproduce the exponential growth dynamics compatible with the trend of the variance evolution of the mother cell size variance. Note that we re-scaled forward scattering intensities by the mean of the starting population to work with smaller numbers. Figure~\ref{fig:5} shows the results of the best fit between the experimental data and the model. In particular, performed a standard $\chi^2$ minimization of the squared residues of the fractions of cells in the different generations (Figure~\ref{fig:5}a), the asymptotic mean and variance of the whole population size (Figure~\ref{fig:5}b1-b2). We opted to leave the mean size per generation out of the scoring function to act as an independent validation of the model. As one can see from  Figure~\ref{fig:5}c, the parameters of the best fit perfectly reproduce also the $<s>_g(t)$ trends (as also testified by the values of the $R^2$). Notably, the optimal value of the $\beta$ exponent of the cell division rate is equal to 6, which indicates a near-adder strategy for size homeostasis (see Figure~\ref{fig:5}d). Note that in the proposed model, a $\beta$ of 1 would have indicated a perfect adder strategy, while $\beta\to\infty$ is a perfect size sensing.

\section{Discussions}


Cell size is a phenotype that exhibits a huge variability across different kinds of cells. Typical sizes of bacteria span a range of 1-10~$\mu m$, eukaryotic cells have linear sizes of 5-100~$\mu m$, to end at neuronal cells whose size is up to some meters.
Besides such inter-kind size heterogeneity, cells of isogenic populations have a well-defined typical size~\cite{Levin2015}, that influences and is influenced in return by processes such as transcription, translation, and metabolism~\cite{DePaiva2005, Miettinen2014}, as cell volume and surface area affects molecule reactions and nutrient exchanges~\cite{Marshall2012}.
How this typical size is preserved despite the complex and noisy machinery~\cite{Pancaldi2014, Miotto2023} of cellular processes that are at play in proliferating cells, is a question that remains still largely unanswered~\cite{Ginzberg2015}, especially for cancer cells that have ten-fold slower proliferation timescales than bacteria or yeast cells.  

Over the years, several models have been proposed to explain how cells regulate their size. These models provide insights into the molecular mechanisms that govern cell growth and division and help researchers identify key regulatory pathways that could be targeted for therapeutic purposes.
In particular, (i) cells that divide after a certain time from birth are said to follow a timer process; (ii) if division takes place when the cell reaches a certain size one speak of sizer model; while (iii) an adder mechanism consists in adding a certain volume which does not depend on the birth size.   

 The `canonical' way to assess the size homeostatic strategy adopted by a certain cell type is based on the trend shown by the size at birth vs that at division.  To obtain such a relation, one has to follow the proliferation of single cells and measure the size of the same cell at its birth and just after its entry into the mitotic phase. While such a procedure provides a reliable way to determine the homeostatic behavior and a sure way to track cell lineages, it also has the limitation of measuring the birth and division sizes, i.e. following the dynamics of individual cells. 

To address such problems, we sought an alternative/complementary procedure able to provide high statistics while preserving cell growth conditions.   We propose an experimental protocol that does not explicitly consider birth and division size but instead utilizes flow cytometry data in combination with a minimal mathematical model to determine the growth and division mechanism of Jurkat cancer T-cells.

Our main finding is that a model based on power functions of the size for division and growth rates successfully reproduces the key features of cell size dynamics, including an Erlang distribution of division times (tasks) and a size-like strategy.
Interestingly, our results are in accordance with those of Tzur \text{et al.}~\cite{Tzur2009} who showed that growth rate is size-dependent throughout the cell cycle in lymphoblasts.

Analyzing the maximum fraction of cells found in the first generation, we found that for the model to correctly reproduce observations, cell division time has to be given by the sum of a minimum of three independent and exponentially distributed intermediate times. 
Notably, this is in accordance with the evidence Chao and coworkers provide of the human cell cycle as a series of uncoupled, memory-less phases~\cite{Chao2019}.
In particular, it is well known that the cell cycle is canonically described as a series of four consecutive phases: G1, S, G2, and M.  Using time-lapse microscopy, Chao \textit{et al.}~\cite{Chao2019} found that each phase duration follows an Erlang distribution and is statistically independent from other phases. 
Interestingly, we found that subpopulation distributions are well described by  Gaussians, thus a good description of the size dynamics is provided following the evolution of first and second moments. 
Finally,  we measure the division noise as CTV noise through correlation analysis, finding that the shapes of the generation fraction curves were compatible with a symmetrical division with fluctuations around the mean up to ten percent.

The model we present is not limited to exponential growth, a major assumption in most of the analytical modelizations present in literature~\cite{Jia2021}. In fact, while this common assumption holds for various cell types,  it is not universal. For example, the evolution of the size distributions in Schizosaccharomyces pombe (fission yeast) is a case where the increase of cell size with time after birth is non-exponential \cite{Nobs2014, Nakaoka2017}. 

We note that our results depend on both the generation fractions and the FSC signal. While fraction signals are reasonably solid,  the forward scattering can only be considered as a proxy for cell size, thus future works should focus on finding a better descriptor. From the analytical point of view, further investigations on both fluctuations on the key parameters (e.g. $\kappa$  and $\lambda$), and on combining different strategies in different tasks should be performed.

In conclusion, we proposed an experimental and theoretical apparatus to characterize the growth and division of leukemia cells.
We found that (i) while following a size-like homeostatic strategy, Jurkat cells (ii) need to pass a certain number of intermediate states before dividing, which are independent and exponentially distributed.  (iii) Experimental data are well reproduced by a minimal model that depends on relatively few, physically meaningful parameters.

\section{Materials and Methods}

\subsection*{Cell culture} 

E6.1 Jurkat cells (kindly provided by Dr. Nadia Peragine, Department of Cellular Biotechnologies and Hematology, Sapienza University of Rome) were used as a cell model for proliferation study and maintained in RPMI-1640 complete culture media containing 10\% FBS, penicillin/streptomycin plus glutamine at 37 C in 5\% CO2. Upon thawing, cells were passaged once prior to amplification for the experiment. Cells were then harvested, counted, and washed twice in serum-free solutions and re-suspended in PBS for further staining. 

\subsection*{Cells fluorescent dye labeling}

To track cell proliferation by dye dilution establishing the daughter progeny of a completed cell cycle, cells were stained with CellTrace\textsuperscript{\texttrademark} Violet stain (CTV, C34557, Life Technologies, Paisley, UK), typically used to monitor multiple cell generations, and MitoTracker\textsuperscript{\textregistered} Deep Red 633 (M22426, Molecular Probes, Eugene, USA). To determine cell viability, prior to dye staining, the collected cells were counted with the hemocytometer using the dye exclusion test of Trypan Blue solution, an impermeable dye not taken up by viable cells. To reduce the time that cells are in incubation with different dyes, we optimized the protocol performing the simultaneous staining of CTV and Mitotracker Deep Red.
For the dyes co-staining, highly viable  $20\times10^6$ cells were incubated in a 2ml solution of PBS containing both CTV (1/1000 dilution according to the manufacturer’s instruction) and Mitotracker Deep Red (used at a final concentration of 200nM) for 25 min at room temperature (RT) mixing every 10min to ensure homogeneous cell labeling. Afterward, complete media was added to the cell suspension for an additional 5 min incubation before the final washing in PBS.

\subsection*{Cell sorting}

Jurkat cells labeled with dyes were sorted using a FACSAriaIII (Becton Dickinson, BD Biosciences, USA) equipped with Near UV 375nm, 488nm, 561nm, and 633nm lasers and FACSDiva software (BD Biosciences version 6.1.3). Data were analyzed using FlowJo software (Tree Star, version 9.3.2 and 10.7.1). Briefly, cells were first gated on single cells, by doublets exclusion with morphology parameters, both side and forward scatter, area versus width (A versus W). The unstained sample was used to set the background fluorescence for each channel. For each fluorochrome, a sorting gate was set around the max peak of fluorescence of the dye distribution~\cite{Filby2015}. In this way, the collected cells were enriched for the highest fluorescence intensity for the markers used. Following isolation, an aliquot of the sorted cells was analyzed with the same instrument to determine the post-sorting purity and population width, resulting in an enrichment $>99$ \% for each sample.

\subsection*{Time course kinetic for dye dilution assessment}
The sorted cell population was seeded into a single well of a 6-well plate (BD Falcon) at 1x10*6 cells/well and kept in culture for up to 72 hours. To monitor multiple cell division, an aliquot of the cells in culture was analyzed every 18, 24, 36, 48, 60, and 72 hours for the fluorescence intensity of CTV dye by the LSRFortessa flow cytometer. To set the time zero of the kinetic, prior culturing, a tiny aliquot of the collected cells was analyzed immediately after sorting at the flow cytometer. The unstained sample was used to set the background fluorescence as described above. Every time that an aliquot of cells was collected for analysis, the same volume of fresh media was replaced in the culture.

\subsection*{Expectation-Maximization and the Gaussian Mixture Model}

We used the Expectation-Maximization (EM) algorithm
to detect the clusters
in Gaussian Mixture Models~\cite{Dempster1977}. The EM algorithm is composed of two steps the
Expectation (E) step and the Maximization (M) step.
In the E-step, for each data point $\bf{f}$,
we used our current guess of $\pi_g $, $\mu_g$, and $\sigma_g$,
to estimate the posterior probability that each cell belongs to generation $g$
given that its fluorescence intensity measure
as $\bf{f}$, $\gamma_g=\mathrm{P}(g|\bf{f})$.
In the M-step, we use the fact that the gradient of the log-likelihood of 
$p(\bf{f}_i)$ for $\pi_g $, $\mu_g$, and $\sigma_g$
can be computed. Consequently, the expression of
the optimal value of $\pi_g $, $\mu_g$, and $\sigma_g$ is dependent on $\gamma_g$.
It is shown that under, certain smoothness conditions the iterative computation
of E-stem and M-step leads us to the locally optimal estimate of the parameters
$\pi_g $, $\mu_g$, and $\sigma_g$, and returns
the posterior probability $\gamma_g$ which weights how much each point belongs to one of the clusters.
Here, we used this model
to perform cluster analysis and detect the peaks which correspond to different
generations. Then, we estimated $\pi_g$, $\mathrm{E}[f_g]$,
and $\mathrm{Var}[f_g]$ from these clusters.

\subsection*{Gillespie simulation}

To validate the mathematical model that was formulated, stochastic simulations of the
growth and dividing cell population were carried out. Note that, through simulation, we can also know the birth and division sizes and can therefore compare the trends of  $\Delta$  Vs $<s_b>$.

In particular, simulations were performed starting from $N=1000$ initial cells, having initial size randomly sampled from a normal distribution of mean $\mu_s$ and variance $\sigma^2_s$.

For each cell, a division time is extracted from the probability distribution $P(t_d)$ via inverse transform sampling.  For the considered system, $P(t_d)$ is given by~\cite{NietoAcuna2019}:
\beq
P(t_d) = 1 - \exp \Big( - \int_0^{t_d} dt h(s)\Big)
\eeq

Upon division, each cell is split into two new daughter cells, each inheriting a fraction $p$ and $(1-p)$ of the mother size, respectively.


\section*{Data Availability}

The data that support the findings of this study are available from the corresponding
author upon reasonable request.

\section*{Code Availability}

All codes used to produce the findings of this study are available from the corresponding author upon request.
The code for the Gaussian Mixture algorithm is available at \href{https://github.com/ggosti/fcMGM}{https://github.com/ggosti/fcGMM}.\\

\section*{Author contributions statement}

M.M., G.G., and G.P. conceived research;  M.L. and G.R. contributed additional ideas; G.P. and S.S. performed experiments;  M.M., S.S., and G.G.  analyzed data; M.M. performed analytical calculations, numerical simulations, and statistical analysis; all authors analyzed results; M.M. wrote the paper; all authors revised the paper.

\section*{Competing Interests}
The authors declare no competing interests.

\section*{Acknowledgements}
G.R. thanks the European Research Council  Synergy grant ASTRA (n. 855923) and the European Innovation Council through its Pathfinder Open Programme, project ivBM-4PAP (grant agreement No 101098989)for support.
M.L. and G.G. thank Project LOCALSCENT, Grant PROT. A0375-2020- 36549, Call POR-FESR “Gruppi di Ricerca 2020”.

\appendix

\section{Derivation of the population size equation}
\label{app:PBE}
To derive the population size equation in Eq.~\ref{eq:PBE}, we start  considering the general case of the phase space spanned by state vectors, $\vec{x}=(x_1, x_2.., x_N)$. An infinitesimal volume in this N-dimensional space is $\delta V=\delta x_1\delta x_2\cdot...\cdot \delta x_N$ so that the area of each hyper-plane is  $\delta A = \frac{\delta V}{\delta x_i}$. Next, we introduce the number of particles in the volume as $\delta N(\vec{x}) = n(\vec{x}) \delta V$, where $n(\vec{x})$ is the particle density. 
Imposing the general balance condition, the accumulation of particles in each volume must equal the flow of particles entering that volume, plus the ones that are generated (e.g. by cell division), minus the ones exiting the volume:

\beq
\label{eq:PBE_1}
\frac{\partial}{\partial t} (\delta N) = - \sum_i^N \delta (n u_i)\delta A_i + \mathcal{G}\delta V
\eeq

In particular, the accumulation on the left is given by the rate of change of the number of particles in the control volume, $\frac{\partial}{\partial t} (\delta N) = \frac{\partial}{\partial t} ( n(\vec{x}) \delta V)$. The first term on the right side instead models the total net flow of particles in the control volume. In fact, it is the sum of the difference between in and out flows of particles across all hyper-planes. The in flows are given by $n~\vec u \cdot (\delta A_i \hat e_i) = n~u_i \delta A_i$, which is the scalar product of the flux and hyperplane perpendicular to the flow.  

Note that $\vec u$ is the velocity vector, while $\hat e_i$ is a versor parallel to the $x_i$ component of the state vector. 

The out flow can be defined as $(n~u_i + \delta (n~u_i))\delta A_i$. Finally, $\mathcal{G}$ contains all production/degradation processes taking place inside the infinitesimal volume.

Substituting all terms in Eq.~\ref{eq:PBE_1}, one obtains
\beq
\label{eq:PBE_2}
\frac{\partial}{\partial t} (n(\vec{x})\delta V) = - \sum_i^N \delta (n~u_i) \frac{\delta V}{\delta x_i} + \mathcal{G}\delta V
\eeq

Dividing each term by $\delta V$ and taking the limit $\delta x \rightarrow 0$, we get the final form:

\beq
\label{eq:PBE_3}
\frac{\partial n }{\partial t} + \vec\nabla \cdot  (\vec{u}n ) - \mathcal{G} = 0
\eeq

\section{Normal moment closure}
\label{app:moment_closure}

The moment's equations we got are mutually dependent and in general not closed~\cite{Totis2021}. To obtain a closed set, we impose the simplest moment closure scheme, i.e. we assume that the distributions are normal. The normal distribution has the property that all odd central moments higher than the first (the mean) are  nil while the even ones are given by:

\beq
\braket{(s - \braket{s})^p} = \begin{cases} 
0 \quad \text{p is odd}\\
\sigma^p(p-1)!!  \quad \text{p is even}\\
\end{cases}
\eeq 

So, for instance, 

\beq
\braket{(s - \braket{s})^3} = \braket{s^3} -3 \braket{s}\braket{s^2} + 2\braket{s}^3 = 0
\eeq

\beq
\braket{s^3} = 3 \braket{s}\braket{s^2} - 2\braket{s}^3 
\eeq

and 

\beq
\braket{s^4} = 4 \braket{s}\braket{s^3} - 6\braket{s}^2\braket{s^2} + 3\braket{s}^4 - 3(\braket{2}^2 - \braket{s}^2)^2
\eeq

\section{Grow and division with intermediate states} 
\label{app:checkpoints}

The model derived in the main text produces an exponential distribution of the division times~\cite{NietoAcuna2019}.
To retrieve an Erlang-like distribution, we introduce a series of intermediate states, the cell has to go through, before starting the division. Each of these states has an exponential distribution of times.

To do so, we can modify the population balance equation by adding a second index that accounts for the states of the cell.  $\rho_{g,q}$ now indicates the density of cells that divided $g$ times and passed the $q$-th state.

The population balance equation becomes

\begin{multline}
\displaystyle
\frac{\partial}{\partial t} n_{g,q}(s,t) + \frac{\partial}{\partial s}\left(\Lambda n_{g,q}(s,t)\right) = 
-\gamma(s) n_{g,q}(s,t) + \\ + \begin{cases}  \int_0^{\infty} d\eta~ \gamma(\eta)~\delta(s -\eta)~n_{g,q-1}(\eta,t) & \text{if q $>$ 0}\\
 &\\
 2 \int_0^{\infty} d\eta~ \gamma(\eta)~\phi(s|p\eta)~n_{g-1,0}(\eta,t) & \text{if q $=$ 0.}\\
\end{cases}    
\end{multline}

Looking at the variation of the total number of cells at generation $g$ and state $q$, we have:

\beq
\dot N_{g,q} = \frac{d}{dt} N_{g,q}(t) =  \int_0^\infty \frac{\partial}{\partial t} n_{g,q}(s,t) ds 
\eeq

which yields
\beq
\label{eq:ndot_gq}
\frac{\dot N_{g,q}}{N_{g,q}} = - k \braket{s^\beta}_{g,q}  +  k \braket{s^\beta}_{g,q-1}  \frac{N_{g,q-1}}{N_{g,q}} = \Phi_{g,q} \quad\text{if}\quad q > 0
\eeq

or

\beq
\label{eq:ndot_gq}
\frac{\dot N_{g,0}}{N_{g,0}} = - k \braket{s^\beta}_{g,0}  +  2k \braket{s^\beta}_{g-1,0}  \frac{N_{g-1,0}}{N_{g,0}} = \Phi_{g,0} \quad\text{if}\quad q = 0
\eeq

Similarly, we define the fraction of cells belonging to the generation, $g$, and state $q$ as:

\beq
P_{g,q}(t) = \frac{N_{g,q}(t)}{\sum_{h,w} N_{h,w}(t) }
\eeq

Eq.~\ref{eq:ndot_gq} can be used to compute the dynamics of the fractions, $P_{g,q}$, since

\begin{multline}
\label{eq:full_fractions}
\dot{P_{g,q}} = \left( \dot{ \frac{N_{g,q}(t)}{\sum_{h,w} N_{h,w}(t) }} \right) = \\ =  \frac{ \dot{N_{g,q}} \left( \sum_{h,w} N_{h,w}\right) - N_{g,q} \sum_{h,w} \dot{N_{h,w}}   }{\left(\sum_{h,w} N_{h,w}\right)^2} = \\= \frac{\dot{N_{g,q}}}{\sum_{h,w} N_{h,w}} - P_{g,q}   \frac{\sum_{h,w} \dot{N_{h,w}}}{\sum_{h,w} N_{h,w}} ~.   
\end{multline}
 
 Using Eq.~\ref{eq:ndot_gq}, we can explicit $\dot{N_{g,q}}$ and obtain the equations reported in the main text:
 
\begin{multline}
\dot{P_{g,q}} =  -k\braket{s^\beta}_{g,q} P_{g,q} + 2k\braket{s^\beta}_{g,q-1} P_{g,q-1} + \\ + P_{g,q} \sum_{h,w}  \left( k\braket{s^\beta}_{h,w} P_{h,w} - 2^{\delta_{w,0}}k\braket{s^\beta}_{h,w -1} P_{h,w -1} \right)
\end{multline}

for  $q > 0$, while 

\begin{multline}
\dot{P_{g,0}} =  -k\braket{s^\beta}_{g,0} P_{g,0} + k\braket{s^\beta}_{g-1,0} P_{g-1,0} + \\ +  P_{g,0} \sum_{h,w}  \left( k\braket{s^\beta}_{h,w} P_{h,w} - 2^{\delta_{w,0}}k\braket{s^\beta}_{h,w -1} P_{h,w -1} \right)~.  \end{multline}


To obtain expressions for the moment's dynamics, we can start from the probability of finding a cell with size $s$ at time $t$ and generation $g$ and at checkpoint $q$ as

\beq
\rho_{g,q}(s,t) = \frac{n_{g,q}(s,t)}{N_{g,q}(t)}
\eeq

and

\beq
\dot{n_{g,q}} = \dot{N_{g,q}} \rho + N_{g,q} \dot{\rho_{g,q}} 
\eeq

With a few calculations, one gets to the distribution moments evolution equations:

\begin{multline}
\braket{\dot{s}^i}_{g,q} =  \lambda \cdot i\cdot \braket{s^{(\alpha + i -1)}}_{g,q} - \Phi_{g,q}\braket{s^i}_{g,q}  -k\braket{s^{(\beta + i)}}_{g,q} +  \\ + 
k\left< 2~p^i\right>_\pi^{\delta_{q,0}} \braket{s^{(\beta+ i)}}_{g,q -1}  \frac{P_{g,q -1}}{P_{g,q}}
\end{multline}

\section{Division noise}
\label{app:noise}
The last term of Eq.~\ref{eq:moments} is obtained 
as following:
\begin{multline}
    2 \int ds s^i \int d\eta \gamma(\eta) K(s|p\eta) \rho(\eta) = \\ =  2\int ds~s^i  \int d\eta \gamma(\eta) \int_0^1 dp~\pi(p)\delta(s-p\eta)  \rho(\eta) = \\ =
    2\int ds~s^i  \int \frac{df}{p} \gamma(f/p) \int_0^1 dp~\pi(p)\delta(s-f)  \rho(f/p) = \\ =  2\int_0^1 dp~ \frac{\pi(p)}{p}\int~ds~s^i \gamma(s/p)   \rho(s/p) = \\ =  2 <s^i \gamma(s)>_\rho \int_0^1 dp~ \frac{\pi(p)}{p} p^(i+1) = \\ =  2 <s^i \gamma(s)>_\rho <p^i>_\pi 
\end{multline}

where we performed two changes of variables, i.e. $p\eta = f$ and $s/p= v$.

\section{Derivation of size-homeostasis strategies}
\label{app:strategies}
We start considering the two observable commonly measured in mother machine experiments, i.e. the division and birth sizes, $s_d$ and $s_b$, respectively.

In particular, the average size at division  can  be expressed as

\begin{equation}
\label{eq:sd}
\braket{s_d} = \int ds_b ds_d s_d \rho(s_b,s_d) = \int ds_b \rho(s_b) \int_{s_b}^\infty s_d \rho(s_d|s_b) d s_d
\end{equation}

where $\rho(s_b, s_d)$ is the distribution of the birth and division sizes and we make use of the chain rule of the conditional probability of having a size of division $s_d$ given that the size at birth was $s_b$, $\rho(s_d| s_b)$.

Using probability conservation, we have that

\begin{equation}
\rho (s_d | s_b) d s_d = \rho_d(t |s_b) dt    
\end{equation}

and

\begin{equation}
\rho (s_d)  = \frac{\rho(t)}{\frac{ds_d}{dt}} = \frac{\rho_d(t_d)}{g(s_d)}
\end{equation}
   
  where $\rho_d(t |s_b)$ is the probability density function of a cell to divide at the time, $t$ for a cell of initial size $s_b$; and $\frac{ds}{dt} = g(s)$ is the definition of growth rate.

  This in turn is given by
  
  \begin{equation}
      \rho_d(t|s_b) = \frac{d P_d(t|s_b)}{dt} = \frac{d}{dt}  \left( 1 - e^{-\int h(s(t')) dt'}\right)
  \end{equation}
  
  where $h$ is the rate of division and the associated probability can be obtained as one minus the probability of not dividing, which evolves in time as:
  \beq
 \frac{dP_0}{dt} = -h(t) P_0\quad  \text{with}\quad P_0(t) = e^{-\int_0^t h dt'}
  \eeq

Let us now assume that growth and division rates are of the form $g(s) = \lambda s^\alpha$ and $h(s) = \kappa s^\beta$, respectively (see for instance~\cite{Totis2021, NietoAcuna2019}).

Thus, one has

\begin{equation}
\label{eq:rho_d}
\rho(s_d) = \exp \left(-\int_{s_b}^{s_d} \frac{\kappa}{\lambda} s^{\beta-\alpha} ds \right) \frac{\kappa}{\lambda} s_d^{\beta -\alpha}
\end{equation}

We can see from Eq.~\ref{eq:rho_d} that the division size distribution depends only on the ratio between the rate coefficients and the difference of their exponents.  Eq.~\ref{eq:sd} can be analytically solved and it yields different scenarios depending on the value of the exponent difference. 

\subsubsection{Adder-like behaviour}

To begin with, we start considering the simpler case, i.e. the one when the two exponents are equal ($\beta -\alpha = 0$). In this case,  Eq.~\ref{eq:sd} becomes:

\beq
\braket{s_d} =  \int ds_b \rho(s_b) \int_{s_b}^\infty s_d 
\exp\left( -\frac{\kappa}{\lambda} (s_d - s_b)  \right) \frac{\kappa}{\lambda}  
d s_d = 
\eeq

\beq
=  \int ds_b \rho(s_b) ( s_b +  \frac{\lambda}{\kappa} )=  \braket{s_b} + \frac{\lambda}{\kappa}
\eeq

Thus, we ended up with:
\beq
\Delta = \braket{s_d} -\braket{s_b} = \frac{\lambda}{\kappa}
\eeq

which is the constant trend one finds in an adder-like growth and division regime.

\subsubsection{Timer-like regime}

One can repeat the same calculations considering the case in which $\beta -\alpha < 0$, that is to say when the exponent associated with growth is bigger than the one regulating division.
In particular, if we consider $\beta -\alpha = -1$, Eq.~\ref{eq:sd} becomes

\beq
\braket{s_d} =  \int ds_b \rho(s_b)  \int_{s_b}^\infty  
\exp\left( - \frac{\kappa}{\lambda} \ln \left( \frac{s_d}{s_b}\right)  \right) \frac{\kappa}{\lambda}  
d s_d = 
\eeq

\beq
= \int ds_b \rho(s_b)   s_b^{\frac{\kappa}{\lambda}} \int_{s_b}^\infty  
\left( \frac{1}{s_d}\right)^{\frac{\kappa}{\lambda}}  \frac{\kappa}{\lambda}  
d s_d = \frac{\kappa}{\kappa -\lambda } \braket{s_b}
\eeq

and 

\beq
\Delta = \frac{\lambda}{\kappa -\lambda}\braket{s_b}
\eeq

The pure timer model is obtained when $\kappa=2\lambda$.

%

\subsubsection{Sizer-like regime}

Finally, if we consider $\beta -\alpha > 0$, Eq.~\ref{eq:sd} becomes
\beq
\braket{s_d} =  \int ds_b \rho(s_b)  \int_{s_b}^\infty  ds_d s_d
\exp \left(-\int_{s_b}^{s_d} \frac{\kappa}{\lambda} s^{\beta-\alpha} ds \right) \frac{\kappa}{\lambda} s_d^{\beta -\alpha}
\eeq

\beq
\braket{s_d} \simeq  \sqrt{\frac{\pi}{\frac{\alpha}{2\beta}}} - \braket{s_b}  
\eeq

that displays the negative correlation associated with the sizer division strategy.


\end{document}